\documentclass[onecolumn]{revtex4}

\usepackage{graphicx}
\usepackage{dcolumn}
\usepackage{bm}

\input epsf
\begin{document}

\title{\Large{Statefinder Parameters for Different Dark Energy Models with Variable
 $G$ Correction in Kaluza-Klein Cosmology}}

\author{\bf Shuvendu Chakraborty$^1$\footnote{shuvendu.chakraborty@gmail.com},
Ujjal Debnath$^2$\footnote{ujjaldebnath@yahoo.com,
ujjal@iucaa.ernet.in},  Mubasher
Jamil$^3$\footnote{mjamil@camp.nust.edu.pk} and Ratbay
Myrzakulov$^{4,5}$\footnote{rmyrzakulov@csufresno.edu,
rmyrzakulov@gmail.com}} \affiliation{$^1$Department of
Mathematics, Seacom Engineering
College, Howrah, 711 302, India.\\
$^2$Department of Mathematics, Bengal Engineering
and Science University, Shibpur, Howrah-711 103, India.\\
$^3$Center for Advanced Mathematics and Physics (CAMP), National
University of Sciences and Technology (NUST), H-12, Islamabad,
Pakistan.\\
$^4$Eurasian International Center for Theoretical
Physics, Eurasian National University, Astana 010008, Kazakhstan.\\
$^5$Department of Physics, California State University, Fresno,
CA 93740 USA.}

\begin{abstract}
In this work, we have calculated the deceleration parameter,
statefinder parameters and EoS parameters for different dark
energy models with variable $G$ correction in homogeneous,
isotropic and non-flat universe for Kaluza-Klein Cosmology. The
statefinder parameters have been obtained in terms of some
observable parameters like dimensionless density parameter, EoS
parameter and Hubble parameter for holographic dark energy, new
agegraphic dark energy and generalized Chaplygin gas models.
\end{abstract}

\maketitle

\sloppy \tableofcontents

\section{Introduction}

Recent cosmological observations obtained by SNe Ia \cite{1}, WMAP
\cite{2}, SDSS \cite{3} and X-ray \cite{4} indicate that the
observable universe experiences an accelerated expansion. To explain
this
 phenomena the notion  known as dark energy (DE) with large negative pressure is proposed.
 At present there are a lot of theoretical models of DE. But the most suitable models  of
 DE is the cosmological constant. According of the modern observational cosmology,
 the present value of cosmological constant is $10^{-55} cm^{-2}$. At the same time,
 the  particle physics tells us that  its value must be $10^{120}$ times greater than
  this factor. It is one main  problem modern cosmology and  known as the cosmological
  constant problem. In order to solve this problem, some authors  considered the cosmological
  constant as a varying parameter (see e.g.  \cite{5,6,7,8,9}).
Here we can mention that Dirac showed that some fundamental
constants do not remain constant forever rather they vary with time
due to some causal connection between micro and macro physics
\cite{10} that is known as Large Number Hypothesis (LNH). The  field
equations of General Relativity (GR) involve two physical constants,
namely, the gravitational constant $G$ (couples the geometry and
matter) and cosmological constant $\Lambda$ (vacuum energy in
space). According to the LNH, the gravitational constant should also
vary with time. In  \cite{11}  LNH was extended by taking
cosmological constant as $\Lambda=\frac{8\pi G^2m^2_p}{h^4}$, where
$m^2_p$ is the mass of proton
 and $h$ is the Planck's constant. It was  showed that $\Lambda$ produces the same
 gravitational effects in vacuum as that produced by matter \cite{11}. As result, this
  cosmological term must be included in the physical part of the field equations.
  In \cite{11} also defined gravitational energy of the vacuum as the interactions of
  virtual particles separated by a distance $\frac{h}{m_p c}$, where $c$ is the speed of
  light. It is also interesting to note that a time varying
gravitational constant also appears in the entropic interpretations
of gravity \cite{momi}.

In the literature, many modifications of cosmological constant
have been proposed for the
 better description and understanding of DE (see e.g. \cite{12}). For example, in  \cite{13} was studied
 the field equations by using three different forms of the cosmological constant,
  i.e., $\Lambda\sim\left(\frac{\dot{a}}{a}\right)^2$, $\Lambda\sim\left(\frac{\ddot{a}}{a}\right)$
  and $\Lambda\sim\rho$ and  shown that these models yield equivalent results to the FRW spacetime.
  From these investigations follow that  an investigation about the scale factor and other
   cosmological parameters with varying $G$ and $\Lambda$ may be interesting especially for
   description the accelerated expansion of the universe.

According modern point of views, multidimensional gravity
theories may play important role to explain main problems of
cosmology and astrophysics in particular DE. One of classical
examples
 of such theories is the theory of Kaluza--Klein (KK) \cite{14,15}. It is a 5 dimensional GR
 in which extra dimension is used to couple the gravity and electromagnetism
 (see e.g., the review \cite{16,17,18} and references therein). In the context of our interest - DE,
 recently  it  was  studied \cite{19} that the non-compact, non-Ricci KK theory and coupled the flat
 universe with non-vacuum states of the scalar field. For  the suitable choose of the equation
 of state (EoS), the reduced field equations  describe the early inflation and late time
 acceleration. Moreover, the role played by the scalar field along the 5th
coordinate in the 5D metric is in general very impressed by the
role of scale factor over the 4D universe.

In recent years, the holographic dark energy (HDE) has been
studied as a possible candidate for DE. It is motivated from the
holographic principle which might lead to the quantum gravity to
explain the events involving high energy scale.  Another
interesting models of DE are the so-called new-agegraphic dark
energy which is originated from the uncertainty  relation of
quantum mechanics together with the gravitational effect of GR. In
general, the agegraphic  DE model assumes that the observed DE
effect  comes from spacetime and matter field fluctuations in the
universe.

In the interesting paper  \cite{20} it was introduced a new
cosmological diagnostic pair $\{r,s\}$ called statefinder which
allows one to explore the properties of DE independent of model.
This pair depends on the third derivative of the scale factor,
$a(t)$, just like the dependence of Hubble and deceleration
parameter on first and second derivative of respectively. It is
used to distinguish flat models of the DE and  this pair has been
evaluated for different models \cite{21,22,23,24,25,26,27,28,29}.
In \cite{29} it was solved the field equations of the FRW universe
with variable $G$ and $\Lambda$ (see also \cite{30} where was
considered the flat KK universe with variable $\Lambda$ but
keeping $G$ fixed). There are many works on higher dimensional
space-time also \cite{31}.

In this work, we have calculated the statefinder parameters for
different dark energy models with variable $G$ correction in
Kaluza-Klein  cosmology. We evaluate different cosmological
parameters with the assumption that our universe is filled with
different types of matter. The scheme of the paper is as follows.
In the next section, the KK model and its field equations are
presented. In section III, solution of the field equations for the
HDE are presented and section IV  the new-agegraphic dark energy
case is considered. Generalized Chaplygin Gas model is studied in
the section V.  In section VI, we summarize the results.

\section{Kaluza-Klein Model}

The metric of a homogeneous and isotropic universe in the
Kaluza-Klein model is

\begin{equation} ds^{2}= dt^{2}-a^{2}(t)\left[\frac{dr^{2}}{1-kr^{2}}+
r^{2}(d\theta^{2}+\sin^{2}\theta d\phi^{2})+(1-k r^{2})d
\psi^{2}\right]
\end{equation}
where $a(t)$ is the scale factor, $k = -1, 0, 1$ is the curvature
parameter for spatially closed, flat and open universe
respectively.\\

We assume that the universe is filled with dark energy and matter
whose energy-momentum tensor is given by

\begin{equation}
T_{\mu \nu}=(\rho_{m}+\rho_{x}+p_{x})u_{\mu}u_{\nu}-p_{x} g_{\mu
\nu}
\end{equation}
where $u_{\mu}$ is the five velocities satisfying
$u^{\mu}u_{\mu}=1$. $\rho_{m}$ and $\rho_{x}$ are the energy
densities of matter and dark energy respectively and $p_{x}$ is the
pressure of the dark energy. We consider here the pressure
of the matter as zero.\\

The Einstein's field equations are given by
\begin{equation}
R_{\mu \nu}-\frac{1}{2}g_{\mu \nu}R=8 \pi G(t)T_{\mu \nu}
\end{equation}
where $R_{\mu \nu}$, $g_{\mu \nu}$ and $R$ are Ricci tensor,
metric tensor and Ricci scalar respectively. Here we consider
gravitational constant $G$ as a function of cosmic time $t$.
 Now from the equations (1), (2) and (3) we have the Einstein's field
equations for the isotropic Kaluza-Klein space time (1) are

\begin{equation}
H^{2}+\frac{k}{a^{2}}=\frac{4\pi G(t)}{3}( \rho_{m}+\rho_{x})
\end{equation}

\begin{equation}
\dot{H}+2H^{2}+\frac{k}{a^{2}}=-\frac{8\pi G(t)}{3} p_{x}
\end{equation}
Let the dark energy obeying the equation of state $p_{x}=\omega
\rho_{x}$. Equation (4) gives

\begin{equation}
\Omega=\Omega_{m}+\Omega_{x}-\Omega_{k}
\end{equation}
where $\Omega_{m},$ $\Omega_{x}$ and $\Omega_{k}$ are dimensionless
density parameters  representing the contribution in the total
energy density. The deceleration parameter $q$ in terms of these
parameters are given by
\begin{equation}
q=\Omega_{m}+(1+2 \omega)\Omega_{x}     \qquad\qquad \text{where}
\qquad\qquad \omega=\frac{q-\Omega-\Omega_{k}}{2 \Omega_{x}}
\end{equation}

The trajectories in the \{$r,s$\} plane \cite{32} corresponding to
different cosmological models depict qualitatively different
behaviour. The statefinder diagnostic along with future SNAP
observations may perhaps be used to discriminate between different
dark energy models. The above statefinder diagnostic pair for
cosmology are constructed from the scale factor $a$. The
statefinder parameters are given by
$$
r=\frac{\dddot{a}}{aH^{2}}~,~~s=\frac{r-1}{3(q-1/2)}
$$

From the expression of one of the statefinder parameter $r$, we
have a relation between $r$ and $q$ is given by

\begin{equation}
r=q+2q^{2}-\frac{\dot{q}}{H}
\end{equation}
From (7) we have
\begin{equation}
\dot{q}=\dot{\Omega}_{m}+(1+2 \omega)\dot{\Omega}_{x}+2
\dot{\omega}\Omega_{x}
\end{equation}
Also we have
\begin{equation}
\Omega=\frac{\rho}{\rho_{cr}}-\frac{k}{a^{2}H^{2}} ~~~~~~~~~
\text{which gives}~~~~
\dot{\Omega}=\frac{\dot{\rho}}{\rho_{cr}}-\frac{2k
q}{a^{2}H}-\frac{\rho \dot{\rho}_{cr}}{(\rho_{cr})^{2}}
\end{equation}
where
\begin{equation}
\rho_{cr}=\frac{3 H^{2}}{4 \pi G(t)}~~~~~~ \text{which gives after
differentiation}~~ \dot{\rho}_{cr}=\rho_{cr}\left(
2\frac{\dot{H}}{H}-\frac{\dot{G}}{G}\right)
\end{equation}
which implies
\begin{equation}
\dot{\rho}_{cr}=-H \rho_{cr}(2(1+q)+\triangle G)
\end{equation}

where, $\triangle G\equiv \frac{G'}{G}, \dot{G}=H G'$. Now from
equation (10) we have

\begin{equation}
\dot{\Omega}=\frac{\dot{\rho}}{\rho_{cr}}+\Omega_{k}H(2+\triangle
G )+\Omega H (2(1+q)+\triangle G)
\end{equation}
We assume that matter and dark energy are separately conserved. For
 matter, $\dot{\rho}_{m}+4 H \rho_{m}=0$. So from (13)
\begin{equation}
\dot{\Omega}_{m}=\Omega_{m} H (-2+2q+\triangle
G)+\Omega_{k}H(2+\triangle G )
\end{equation}
For dark energy, $\dot{\rho}_{x}+4 H (1+\omega)\rho_{x}=0$. So
from (13)
\begin{equation}
\dot{\Omega}_{x}=\Omega_{x} H (-2-4 \omega+2q+\triangle
G)+\Omega_{k}H(2+\triangle G )
\end{equation}
From (8), (9), (14), (15) we have expression for $r$ and $s$
given by

\begin{equation}
r=3\Omega_{m}+(3+10\omega+8\omega^{2})\Omega_{x}-4(1+\omega)\Omega_{k}-\triangle
G(\Omega_{m}+(1+2\omega)\Omega_{x}+2(1+\omega)\Omega_{k})-\frac{2\dot{\omega}\Omega_{x}}{H}
\end{equation}
\begin{equation}
s=\frac{3\Omega_{m}+(3+10\omega+8\omega^{2})\Omega_{x}-4(1+\omega)\Omega_{k}-\triangle
G(\Omega_{m}+(1+2\omega)\Omega_{x}+2(1+\omega)\Omega_{k})-\frac{2\dot{\omega}\Omega_{x}}{H}-1}{3
(-1/2 + \Omega_{m} + \Omega_{x} + 2 \omega \Omega_{x})}
\end{equation}

\section{Holographic Dark Energy}
To study the dark energy models from the holographic principle it is
important to mention that the number of degrees of freedom is
directly related to the entropy scale with the enclosing area of the
system, not with the volume \cite{33}. Where as Cohen et al
\cite{34} suggest a relation between infrared (IR) and the
ultraviolet (UV) cutoff in such a way that the total energy of the
system with size $L$ must not exceed the mass of the same size black
hole. The density of holographic dark energy is

\begin{equation}
\rho_{x}=\frac{3 c^{2}}{8 \pi G}\frac{1}{L^{2}}
\end{equation}
Here $c$ is the holographic parameter of order unity. Considering
$L=H_{0}^{-1}$ one can found the energy density compatible with
the current observational data. However, if one takes the Hubble
scale as the IR cutoff, the holographic dark energy may not
capable to support an accelerating universe \cite{35}. The first
viable version of holographic dark energy model was proposed by Li
\cite{36}, where the IR length scale is taken as the event horizon
of the universe. The holographic dark energy has been explored
in various gravitational frameworks \cite{setare}\\

The time evolution is
\begin{equation}
\dot{\rho}_{x}=-\rho_{x}H(2-\frac{2\sqrt{2\Omega_{x}}}{c}cos
y+\triangle G)
\end{equation}
where $L$ is defined as $L=a r(t)$ with $a$ is the scale factor.
Also $r(t)$ can be obtained from the
relation $R_{H}=a\int\limits_{t}^{\infty}\frac{d t}{a}=\int_{0}^{r(t)}
\frac{dr}{\sqrt{1-k r^{2}}}$.\\

where $R_{H}$ is the event horizon. When $R_{H}$ is the radial
size of the event horizon measured in the $r$ direction, $L$ is
the radius of the event horizon measured on the sphere of the
horizon.\\

For closed (or open) universe we have $r(t)=\frac{1}{\sqrt{k}}sin
y $, where
$y=\frac{\sqrt{k}R_{H}}{a}$.\\
 using
the definition $\Omega_{x}=\frac{\rho_{x}}{\rho_{cr}}$ and
$\rho_{cr}=\frac{3 H^{2}}{4 \pi G(t)}$ we have
$HL=\frac{c}{\sqrt{2 \Omega_{x}}}$.\\

And using all these we ultimately obtain the relation
$\dot{L}=HL+a\dot{r}(t)=\frac{c}{\sqrt{2 \Omega_{x}}}-cos y$,
by which we find the equation (19).\\

From the energy conservation equation and the equation (19) we
have the holographic energy equation of state given by
\begin{equation}
\omega=\frac{1}{4}\left(-2-\frac{2\sqrt{2\Omega_{x}}}{c}cos
y+\triangle G\right)
\end{equation}

where, $\Omega_{k}=\frac{k}{a^{2}H^{2}}$,
$\Omega_{x}=\frac{c^{2}}{2 L^{2}H^{2}}$ are usual fractional
densities in KK model.\\
 From the ration of the fractional densities we have, $sin^{2}y=\frac{c^{2}\Omega_{k}}{2
\Omega_{x}}$ and naturally,  $cos
y=\sqrt{\frac{2 \Omega_{x}-c^{2}\Omega_{k}}{2 \Omega_{x}}}$.\\

Now differentiating (20) and using (15) we have

\begin{equation}
\frac{\dot{\omega}}{H}=\frac{16 \Omega_{x}^2 (-1 + \Omega_{x}) +
c^2 \Omega_{x} (3 \triangle 'G +\Omega_{k}(2- 8 \Omega_{x})) -
 4c \sqrt{-c^2 \Omega_{k} + 2 \Omega_{x}} ((2 + \triangle  G)
  \Omega_{k} + \Omega_{x} (2 \Omega{m} +
  \triangle  G \Omega_{x}))}{12 c^2 \Omega_{x}}
\end{equation}
Now putting (21) in (16) and (17), we have
\begin{eqnarray*}
r=\frac{1}{6 c^2}\left[8 (5 - 2 \Omega_{x}) \Omega_{x}^2 -
 c^2 (3 (2 (-3 + \triangle  G) \Omega_{m} + (-\triangle G +
 \triangle' G) \Omega_{x}) + \Omega_{k} (3 (2 + \triangle G)^2 + 14 \Omega_{x} - 8
 \Omega_{x}^2))\right.
\end{eqnarray*}

\begin{equation}
\left.+2 c \sqrt{-c^2 \Omega_{k} + 2 \Omega_{x}} (5 (2 +
\triangle G) \Omega_{k} + \Omega_{x} (-3 + 4 \Omega_{m} +
\triangle G(-3 + 2 \Omega_{x})))\right]
\end{equation}

\begin{eqnarray*}
 s=\frac{1}{9 c (-2 \Omega_{x} \sqrt{-c^2 \Omega_{k} + 2 \Omega_{x}} + c (-1 + 2 \Omega_{m} + \triangle G \Omega_{x}))}\left[8 (5 - 2 \Omega_{x}) \Omega_{x}^2 -
 c^2 (3(2+2 (-3 + \triangle G) \Omega_{m} + (-\triangle G + \triangle' G) \Omega_{x}) \right.
\end{eqnarray*}

\begin{equation}
\left.+ \Omega_{k} (3 (2 + \triangle G)^2 + 14 \Omega_{x} - 8
 \Omega_{x}^2))+2 c \sqrt{-c^2 \Omega_{k} + 2 \Omega_{x}} (5 (2 + \triangle
G) \Omega_{k} + \Omega_{x} (-3 + 4 \Omega_{m} + \triangle G (-3 +
2 \Omega_{x})))\right]
\end{equation}
This is the expressions for $\{r,s\}$ parameters in terms of
fractional densities of holographic dark energy model in
Kaluza-klein cosmology for closed (or open) universe.\\

\section{\normalsize\bf{New Agegraphic Dark Energy}}
There are another version of the holographic dark energy model
called, the new agegraphic dark energy model \cite{37}, where the
time scale is chosen to be the conformal time. The new agegraphic
dark energy is more acceptable than the original agegraphic dark
ennergy, where the time scale is choosen to be the age of the
universe. The original ADE suffers from the difficulty to describe
the matter-dominated epoch while the NADE resolved this issue. The
density of new agegraphic dark energy is
\begin{equation}
\rho_{x}=\frac{3 n^{2}}{8 \pi G}\frac{1}{\eta^{2}}
\end{equation}
where $n$ is a constant of order unity. where the conformal time
is given by
$\eta=\int_{0}^{a}\frac{da}{Ha^{2}}$.\\
 If we consider $\eta$ to
be a definite integral, the will be a integral constant and we
have $\dot{\eta}=\frac{1}{a}$.\\
Considering KK cosmology and using the definition
$\Omega_{x}=\frac{\rho_{x}}{\rho_{cr}}$ and $\rho_{cr}=\frac{3
H^{2}}{4 \pi G(t)}$ we have
$H\eta=\frac{n}{\sqrt{2 \Omega_{x}}}$.\\
After introducing the fractional energy densities we have the time
evolution of NADE as
\begin{equation}
\dot{\rho_{x}}=-\rho_{x}H\left(\frac{2\sqrt{2 \Omega_{x}}}{n
a}+\triangle G\right)
\end{equation}
From the energy conservation equation and the equation (25) we
have the new agegraphic energy equation of state given by

\begin{equation}
\omega=\frac{1}{4}\left(-4+ \frac{2\sqrt{2 \Omega_{x}}}{n
a}+\triangle G\right)
\end{equation}

where, $\Omega_{k}=\frac{k}{a^{2}H^{2}}$,
$\Omega_{x}=\frac{n^{2}}{2 \eta^{2}H^{2}}$ are usual fractional
densities in KK model.\\

 Differentiating (26) and using (15) we have

 \begin{equation}
 \frac{\dot{\omega}}{H}=\frac{a^2 \triangle' G n^2 \sqrt{x} + 4 (-1 + \Omega_{x}) \Omega_{x}^{3/2} +
 \sqrt{2} a n ((2 + \triangle G) \Omega_{k} + \Omega_{x} (2 \Omega_{m} + (-2 + \triangle G) \Omega_{x}))}{4 a^2 n^2 \sqrt{\Omega_{x}}}
 \end{equation}
 Now putting (27) in (16) and (17), we have the expression for $r,s$
 as

\begin{eqnarray*}
 r=-\frac{1}{2 a^{2} n^{2}}\left[4 (-3 + \Omega_{x}) \Omega_{x}^2 +
 \sqrt{2} a n \sqrt{\Omega_{x}} (3 (2 + \triangle G) \Omega_{k} + (2 (3 + \Omega_{m} - \Omega_{x}) + \triangle G (-2 + \Omega_{x})) \Omega_{x})\right.
\end{eqnarray*}
\begin{equation}
 \left.+a^2 n^2 (\triangle G^2 \Omega_{k} - 6 \Omega_{m} + (-2 + \triangle' G) \Omega_{x} + \triangle G (2 (\Omega_{k} + \Omega_{m}) + \Omega_{x}))\right]
\end{equation}

\begin{eqnarray*}
 s=-\frac{1}{3 a n (2 \sqrt{2} \Omega_{x}^{3/2} + a n (-1 + 2 \Omega_{m} + (-2 + \triangle G) \Omega_{x}))}\left[4 (-3 + \Omega_{x}) \Omega_{x}^2 +
 \sqrt{2} a n \sqrt{\Omega_{x}} (3 (2 + \triangle G) \Omega_{k} + (2 (3 + \Omega_{m} - \Omega_{x}) \right.
\end{eqnarray*}

\begin{equation}
 \left.+ \triangle G (-2 + \Omega_{x})) \Omega_{x})+a^2 n^2 (2+\triangle G^2 \Omega_{k} - 6 \Omega_{m} + (-2 + \triangle' G) \Omega_{x} + \triangle G (2 (\Omega_{k} + \Omega_{m}) + \Omega_{x}))\right]
\end{equation}
This is the expressions for $\{r,s\}$ parameters in terms of
fractional densities of new agegraphic dark energy model in
Kaluza-klein cosmology for closed (or open) universe.\\

\section{Generalized Chaplygin gas}
It is well known to everyone that Chaplygin gas provides a different
way of evolution of the universe and having behaviour at early time
as presureless dust and as cosmological constant at very late times,
an advantage of  GCG, that is it unifies dark energy and matter into
a single equation of state. This model can be obtained from
generalized version of the Born-Infeld action. The equation of state
for generalized Chaplygin gas is \cite{38}
\begin{equation}
p_{x}=-\frac{A}{\rho_{x}^{\alpha}}
\end{equation}
 where $0<\alpha<1$ and $A>0$ are constants. Inserting the above equation
  of state (30) of the GCG into the energy conservation equation
 we have
\begin{equation}
\rho_{x}=\left[A+\frac{B}{a^{4(\alpha+1)}}\right]^{\frac{1}{\alpha+1}}
\end{equation}
where $B$ is an integrating constant.
\begin{equation}
\omega=-A \left(A + \frac{B}{a^{4 (1 + \alpha)}}\right)^{-1}
\end{equation}
 Differentiating (32) and using (15) we have

\begin{equation}
 \frac{\dot{\omega}}{H}=-4 A B (1 + \alpha) \frac{1}{a^{
 4 (1 + \alpha)}} \left(A + \frac{B}{a^{4 (1 + \alpha)}}\right)^{-2}
\end{equation}
Now putting (33) in (16) and (17), we have

\begin{equation}
r=3 \Omega_{m} - \triangle G \Omega_{m} + \Omega_{x} + \triangle G
\Omega_{x}-\frac{2 B ((2 + \triangle G) \Omega_{k} + \Omega_{x}
(-1 + \triangle G - 4 \alpha))}{\left(a^{4 + 4 \alpha} A +
B\right)} -\frac{8 B^2 \Omega_{x} \alpha}{(A
 a^{4 + 4 \alpha}  + B)^{2}}
 \end{equation}

 \begin{equation}
s=\frac{3 \Omega_{m} - \triangle G \Omega_{m} + \Omega_{x} +
\triangle G \Omega_{x}-\frac{2 B ((2 + \triangle G) \Omega_{k} +
\Omega_{x} (-1 + \triangle G - 4 \alpha))}{\left(a^{4 + 4 \alpha}
A + B\right)} -\frac{8 B^2 \Omega_{x} \alpha}{(A
 a^{4 + 4 \alpha}  + B)^{2}}}{3\left(-1/2+\Omega_{m}+\Omega_{x} -\frac{2 A \Omega_{x}}{A+a^{-4(1+\alpha)}B}\right)}
 \end{equation}
This is the expressions for $\{r,s\}$ parameters in terms of
fractional densities of generalized Chaplygin gas model in
Kaluza-klein cosmology for closed (or open) universe.\\

\section{Conclusions}
In this work, we have considered the homogeneous, isotropic and
non-flat universe in 5D Kaluza-Klein Cosmology. We have calculated
the corrections to statefinder parameters due to variable
gravitational constant in Kaluza-Klein Cosmology. These corrections
are relevant because several astronomical observations provide
constraints on the variability of $G$. We have investigated three
multipromising models of DE such as the Holographic dark energy, the
new-agegraphic dark energy and generalized Chaplygin gas. These dark
energies derive the accelerating phase of the Kaluza-Klein model of
the universe. We have assumed that the dark energies do not interact
with matter. In this case, the deceleration parameter and equation
state parameter for dark energy candidates have been found. The
statefinder parameters have been found in terms of the dimensionless
density parameters as well as EoS parameter $\omega$ and the Hubble
parameter. An important thing to note is that the $G$-corrected
statefinder parameters are still geometrical since the parameter
$\triangle G$ is a pure number and is independent of the geometry.

\subsection*{Acknowledgments}
Special thanks to the referees for numerous comments to improve the
quality of this work.


\begin{thebibliography}{99}
\bibitem{1} Riess A.G. et al.: Astron. J. \textbf{116}(1998)1009;\\
Perlmutter, S. et al.: Astrophys. J. \textbf{517}(1999)565.
\bibitem{2} Tegmark M. et al.: Phys. Rev. \textbf{D69}(2004)103501.
\bibitem{3} Allen S.W. et al.: Mon. Not. Roy. Astron. Soc. \textbf{353}(2004)457.
\bibitem{4} Spergel D.N. et al.: Astrophys. J. Suppl. \textbf{148}(2003)175;\\
Komatsu E. et al.: Astrophys. J. Suppl. 180(2009)330.
\bibitem{5} Ratra B. and Peebles, P.J.E.: Phys. Rev. \textbf{D37}(1988)3406.
\bibitem{6} Dolgov A.D.: Phys. Rev. \textbf{D55}(1997)5881.
\bibitem{7} Sahni V. and Starobinsky, A.: Int. J. Mod. Phys. \textbf{D9}(2000)373.
\bibitem{8} Padmanabhan T.: Phys. Rep. \textbf{380}(2003)235.
\bibitem{9} Peebles P.J.E.: Rev. Mod. Phys. \textbf{75}(2003)599.
\bibitem{10} P.A.M. Dirac, Proc. R. Soc. Lond. A \textbf{165} (1938) 199;\\
A. Beesham, Int.  J. Theor.  Phys. \textbf{33} (1994) 1383;\\
Ray S. et al.: \textit{Large Number Hypothesis},
arXiv:0705.1836v1;\\ M.R. Setare, D. Momeni, Commun. Theor. Phys.
\textbf{56} (2011) 691.
\bibitem{11} Zeldovich Ya.B.: Usp. Nauk. \textbf{95}(1968)209.
\bibitem{momi} D. Momeni , Int. J. Theor. Phys. \textbf{50} (2011) 2582;\\
M.R. Setare, D. Momeni, Commun.Theor.Phys. 56 (2011)
691.
\bibitem{12} Overduin J.M. and Cooperstock, F.I.: Phys. Rev. \textbf{D58}(1998)043506.
\bibitem{13} Ray S. and Mukhopadhyay U.: Grav. Cosmol. \textbf{13} (2007) 142;\\
M.S. Berman, Phys. Rev. Phys. Rev. \textbf{D 43}, 1075 (1991);\\
H. Liu, P. Wesson,  (2001) ApJ \textbf{562} 1;\\ S. Podariu, B.
Ratra, Astrophys. J. \textbf{532} (2000) 109;\\ A. Pradhan, P.
Pandey, Astrophys. Space Sci. \textbf{301} (2006) 127;\\ A.I. Arbab,
Chin. Phys. Lett. \textbf{25} 4497 (2008);\\ A.I. Arbab, Chin. Phys.
Lett. \textbf{25} 3834 (2008)
\bibitem{14} Kaluza T.: Sitz. Press. Akad. Wiss. Phys. Math. \textbf{K1}(1921)966.
\bibitem{15} Klein O.: Zeits. Phys. \textbf{37}(1926)895.
\bibitem{16} Overduin J.M. and Wesson P.S.: Phys. Rept. \textbf{283}(1997)303.
\bibitem{17} Lee H.C.: \textit{An Introdution to Kaluza Klein Theories} (World Scientific, 1984).
\bibitem{18} Appelquist T., Chodos A. and Freund P.G.O.: \textit{Modern Kaluza-Klein Theories} (Addison-Wesley, 1987).
\bibitem{19} Darabi F.: \textit{Dark Pressure in Non-compact and Non-Ricci Flat 5D Kaluza-Klein Cosmology}, arXiv/1101.0666v1.
\bibitem{20} Sahni V. et al.: JETP. Lett. \textbf{77}(2003)201.
\bibitem{21} Zhang X.: Int. J. Mod. Phys. \textbf{D14}(2005)1597.
\bibitem{22} Wei H. and Cai, R.G.: Phys. Lett. \textbf{B655}(2007)1.
\bibitem{23} Zhang X.: Phys. Lett. \textbf{B611}(2005)1.
\bibitem{24} Huang J.Z. et al.: Astrophys. Space Sci. \textbf{315}(2008)175.
\bibitem{25} Zhao W.: Int. J . Mod. Phys. \textbf{D17}(2008)1245.
\bibitem{26} Hu M. and Meng, X.H.: Phys. Lett. \textbf{B635}(2006)186.
\bibitem{27} Zimdahl, W. and Pavon D.: Gen. Relativ. Gravit. \textbf{36}(2004)1483.
\bibitem{28} Shao Y. and Gui Y.: Mod. Phys. Lett. \textbf{A23}(2008)65.

\bibitem{29} Jamil M. and Debnath U.:
Int. J. Theor. Phys. \textbf{ 50} 1602 (2011);\\
Sharif, M., Khanum, F., Astrophys. Space Sci. \textbf{334} 209 (2011);\\
Jamil, M., Int. J. Theor. Phys. \textbf{49} 2829 (2010);\\
M. Jamil, U. Debnath, Astrophys. Space
Sci. \textbf{333} 3 (2011);\\ ibid, Astrophys. Space Sci. \textbf{335} 545 (2011);\\
M.U. Farooq et al, Astrophys. Space Sci. \textbf{334} 243 (2011);\\
Reddy, D.
R. K. and Naidu, R. L., Int. J. Theor. Phys. \textbf{47} 2339 (2008);\\
Darabi, F., Mod. Phys. Lett. A, \textbf{25} 1635 (2010);\\ Darabi,
F., Sajko, W. N. and Wesson, P. S., Class. Quantum Grav. \textbf{17}
4357 (2000).
\bibitem{30} Pradhan A. et al.: Int. J. Theor. Phys.
\textbf{47} (2008) 1751;\\
M. Jamil et al, Eur. Phys. J. C \textbf{60} 149 (2009);\\
 Ozel C., Kayhan H. and Khadekar G.S.: Adv. Studies. Theor. Phys. \textbf{4}(2010)117.
\bibitem{31} R. A. El-Nebulsi, Research in Astron. Astrophys.  \textbf{11} 759
(2011);\\ Tiwari, R. K., Rahaman, F. and Ray, S., Int. J. Theor.
Phys. \textbf{49} 2348 (2010);\\  Farajollahi, H. and Amiri, H.,
Int. J. Mod. Phys. D \textbf{19} 1823 (2010);\\ Huang, B., Li, S.
and Ma, Y., Phys. Rev. D \textbf{81} 064003 (2010);\\ R.A.
El-Nebulsi, Astrophys. Space Sci. \textbf{327}, 111 (2010);\\
Canfora, F., Giacomimi, A. and Zerwekh, A. R., Phys. Rev. D
\textbf{80} 084039 (2009).
\bibitem{32}Alam U. etal.:JETP Lett. \textbf{77} (2003) 201.
\bibitem{33} Susskind L.: J. Math. Phys.\textbf{36} (1995) 6377;\\
't Hooft G: arXiv:9310026 [gr-qc].
\bibitem{34} Cohen A.etal.: Phys. Rev. Lett.\textbf{82} (1999) 4971.
\bibitem{35} S. D. H. Hsu: Phys. Lett. B \textbf{594} (2004) 13.
\bibitem{36} Li M.: Phys. Lett. B \textbf{603} (2004) 1.

\bibitem{setare} M.R. Setare, Phys. Lett. \textbf{B642} (2006) 421;\\
M.R. Setare, Phys. Lett. \textbf{B648} (2007) 329;\\
M. R. Setare, J. Zhang, X. Zhang,     JCAP \textbf{0703} (2007)
007;\\
M. Jamil, M.U. Farooq, M.A. Rashid, Eur. Phys. J. C \textbf{61} 471 (2009);\\
M. Jamil, M.U.Farooq, Int. J. Theor. Phys. \textbf{49} (2010) 42;\\
M.R. Setare, M. Jamil, JCAP \textbf{02} (2010) 010;\\ M. Jamil, M.U.
Farooq, JCAP \textbf{03} (2010) 001;\\ M. Jamil, A. Sheykhi, M.U.
Farooq, Int. J. Mod. Phys. \textbf{D 19} (2010) 1831;\\ H.M.
Sadjadi, M. Jamil, Gen. Rel. Grav. \textbf{ 43} 1759 (2011);\\
M. Jamil et al, Int. J. Theor. Phys, \textbf{51} (2012) 604;\\
M.R. Setare, M. Jamil, Gen. Relativ. Gravit. \textbf{43}, (2011) 293

\bibitem{37}H. Wei and R. G. Cai: Phys. Lett. B \textbf{660} (2008) 113;\\
H. Wei and R. G. Cai, Phys. Lett. B \textbf{663} (2008) 1;\\
Zhang J. etal.:Eur. Phys. J. C \textbf{54} (2008) 303.
\bibitem{38}Gorini V. etal.:Phys. Rev. D \textbf{67} (2003) 063509;\\
Alam U. etal.:Mon. Not. Roy. Astron. Soc. \textbf{344} (2003) 1057;\\
Bento M. C.:Phys. Rev. D \textbf{66} (2002) 043507.




\end{thebibliography}
\end{document}